\documentclass[prd,twocolumn,showpacs,showkeys,amsmath,amssymb]{revtex4}

\usepackage{graphicx}
\usepackage{dcolumn}
\usepackage{bm}
\usepackage[dvips]{color}

\begin{document}

\preprint{APS/123-QED}

\title{Have neutron stars a dark matter core?}

\author{Paolo Ciarcelluti\footnote{Address for correspondence:\\
via Fortore 3, I-65015 Montesilvano, Italy}}
 \email{paolo.ciarcelluti@gmail.com}
 \noaffiliation

\author{Fredrik Sandin}
 \email{fredrik.sandin@gmail.com}
 \affiliation{EISLAB, Lule{\aa} University of Technology, 971 87 Lule{\aa}, Sweden}

\date{\today}

\begin{abstract}
Recent observational results for the masses and radii of some neutron stars are in contrast with typical observations and theoretical predictions for ``normal'' neutron stars.
We propose that their unusual properties can be interpreted as the signature of a dark matter core inside them.
This interpretation requires that the dark matter is made of some form of stable, long-living or in general non-annihilating particles, that can accumulate in the star \cite{Sandin:2008db}.
In the proposed scenario all mass-radius measurements can be explained with one nuclear matter equation of state and a dark core of varying relative size.
This hypothesis will be challenged by forthcoming observations and could eventually be a useful tool for the determination of dark matter.

\end{abstract}




\maketitle


Nowadays dark matter (DM) is an inevitable reality in both astrophysics \cite{Roos:2010wb} and particle physics \cite{Feng:2010gw}.
The evidence for this non-baryonic ingredient of our Universe comes from observations at galactic and super-galactic scales, and it seems clear that it is the largely dominant fraction of matter.
In this scenario it is to be expected that some DM should be present in stellar objects.
Depending on its nature, the mechanisms for accumulating DM are different, and it can result both from the stellar formation process and from subsequent accumulation via capture of DM particles during the whole stellar lifetime.

On this basis, efforts have been made to investigate the possible observational consequences.
To this aim, during the last years attention has been put to DM in relation to neutron stars (NS).
The reasons are that NS should efficiently capture DM due to their high particle density, and new precise results from observations of NS are becoming more frequently available.
Therefore, NS are considered potentially useful objects to help constrain DM models.
In two recent papers \cite{Kouvaris:2010vv,deLavallaz:2010wp} the constraints on the properties of DM, in the form of weakly interacting massive particles (WIMPs), were revised by considering the effects on the luminosity of the accretion and self-annihilation of DM in NS.

Here we follow another approach \cite{Sandin:2008db} and consider the potential effect on the masses and radii of NS equilibrium configurations after accretion of DM.
A key point is that the mass and radius of a NS are independent of the present DM accretion in the star.
Instead, they depend on the whole DM capture process integrated over the stellar lifetime, and effects may be observable even if the star is not presently accreting DM.
For some DM candidates there can be significant effects of DM on the equilibrium sequence of NS, {\it i.e.}, on their masses and radii.
The magnitude of this effect is proportional to the amount of DM trapped in the gravitational field of each NS, which depends on the individual history \cite{Sandin:2008db}, starting from the formation of the progenitor star and continuing through the evolutionary phases until present age.
Our hypothesis predicts significant effects on the mass and radius of some NS, in contrast to the small and presently unobservable modification of the luminosity expected from self-annihilation of WIMPs.
Interestingly, there are now several observations of NS with mass-radius relations that could be in contrast with typical observations and predictions based on nuclear matter models \cite{Ozel:2006km,Ozel:2008kb,Guver:2008gc,Guver:2010td}.
Such observations are used in attempts to constrain the equation of state (EOS) of NS matter \cite{Ozel:2010fw}.
Here we consider observational results for the four NS discussed in those papers.
A combined analysis shows that there is a clear discrepancy between EXO 0748-676 \cite{Ozel:2006km} and the group 4U 1608-52 \cite{Guver:2008gc}, 4U 1820-30 \cite{Guver:2010td} and EXO 1745-248 \cite{Ozel:2008kb}.
In fact, the author of Ref.~\cite{Ozel:2006km} claims that EXO 0748-676 rules out soft EOS for NS, while in the three more recent cases the same author with colleagues conclude that only a soft EOS is consistent with these mass-radius observations.
In order to explain these data it has recently been suggested the existence of two populations of NS, compact and ultra-compact (composed partially or totally of quarks) \cite{Drago:2010rp}, see also \cite{Klahn:2006iw,Alford:2007}.

An alternative explanation, which does not require novel exotic phases of dense matter, is that extraordinary compact NS have a dark matter core.
This is not possible if DM is made of particles similar to the typical theoretical predictions for WIMPs, but only for stable or long-living particles that can accumulate in the centre of NS.
A motivation and a detailed investigation of the effects of one such DM candidate on NS can be found in Ref.~\cite{Sandin:2008db} and references therein.
Here we repeat some of the key points. We consider stable DM particles that interact with ordinary particles by gravity only. 
According to general relativity, all forms of energy are sources of gravity. 
In particular, the curvature of spacetime inside a compact star depends on the energy-density and pressure distribution of matter. 
In presence of DM the metric is affected by both visible and dark components
\begin{eqnarray}
  p(r) &=& p(r) + p_{DM}(r), \label{eq:pressure}\\
  \rho(r) &=& \rho(r) + \rho_{DM}(r). \label{eq:density}
\end{eqnarray}
In the Einstein field equations, the metric functions, $\lambda(r)$ and $\nu(r)$, apply to both baryons and DM, and are modified by the replacements (\ref{eq:pressure}-\ref{eq:density}).
The standard equation for hydrostatic equilibrium separates for the two components
\begin{eqnarray}
	\frac{dp}{dr} &=& -\left[p(r)+\rho(r)\right]\frac{d\nu}{dr}, \label{eq:einstein3O} \\
	\frac{dp_{DM}}{dr} &=& -\left[p_{DM}(r)+\rho_{DM}(r)\right]\frac{d\nu}{dr}, \label{eq:einstein3M}
\end{eqnarray}
because we assume that baryons and DM particles interact by gravity only, {\it i.e.}, the gradient of the pressure of baryons does not exert a direct force on fluid elements of DM, and vice-versa.
This result can be generalized to any number of fluids that interact by gravity \footnote{The effect of eventual weak interactions between visible and DM on the equilibrium structure of a compact star should be small and we therefore consider only the gravitational interaction here.}.

In a previous paper \cite{Sandin:2008db} we have solved these equations and studied the equilibrium properties of neutrons stars that have a core of mirror DM.
This stable and self-interacting DM candidate emerges from the parity symmetric extension of the Standard Model (SM) of particles \cite{Foot:1991bp}.
In this model there exist a hidden set of particles that have the same physical properties as SM particles, but with ``right-handed'' weak interactions.
These ``mirror'' particles interact with baryons essentially by gravity.
For a review see Ref.~\cite{Foot:2004pa} and for cosmological consequences Refs.~\cite{Ciarcelluti:2003wm,Ciarcelluti:2009da}.
Since the microphysics of mirror DM is the same as that of baryons, we use the same nuclear matter EOS for the ordinary and dark matter components of the NS.
In this special case, there are in fact two compact objects in the same location of spacetime.
One is visible and made of ordinary baryons, and the other is hidden and made of mirror baryons.
In this scenario we find that the mass-radius relation is significantly modified in the presence of DM.
This means that the equilibrium sequence of NS is non-unique and history dependent, since it depends on the relative amount of baryons and DM.
Interestingly, the mentioned recent observational results for the masses and radii of NS indicate that there might not exist a unique equilibrium sequence.

\begin{figure}
\includegraphics[scale=1.1]{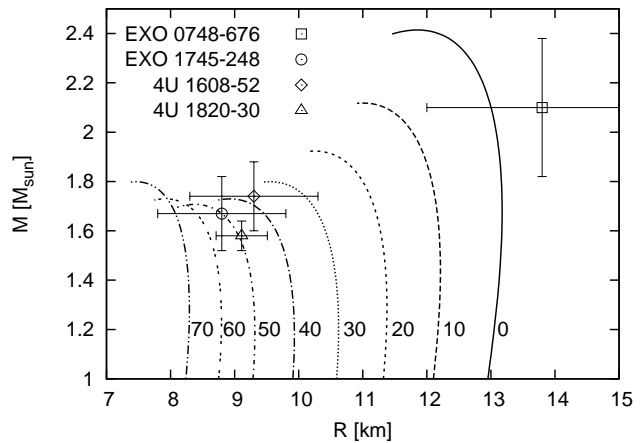}	
\caption{Mass-radius relations for different dark matter (DM) contributions to the total neutron star (NS) mass, computed for the case of mirror DM. The numbers near the lines are mass ratios in percent, {\it i.e.}, the mass of the DM core over the total stellar mass. Numbers larger than 50\% are related to structures where the NS is the core of a bigger compact object made of DM. Observational results for four NS with 1-$\sigma$ errors are included for comparison (see the text for references).}
\label{fig:R-M}
\end{figure}

In Figure \ref{fig:R-M} we plot the mass-radius relations for a moderately stiff EOS (for details see the original paper \cite{Sandin:2008db}) and a varying relative amount of mirror DM, $m_{DM}/m_{tot}$, ranging from 0\% to 70\%.
Values larger than 50\% mean that there is more DM than ordinary matter, and that NS are the cores of bigger dark compact objects.
Such objects should be extraordinary compact, making eventual observations difficult to explain with other methods. 
A feature of this model is the shift of the equilibrium sequence towards lower radii for increasing amounts of DM.
Comparing with current observations, it appears that the less compact NS, EXO 0748-676, is well interpreted with models without DM, while for the other objects considered here we need DM cores of at least 25\% relative mass.
In this way one can explain all the observational results with one nuclear matter EOS and varying amounts of trapped DM.
If this picture is correct, it means that, under some conditions, DM can be efficiently captured by NS, or is present in significant amounts during the formation of some progenitor stars.
In this scenario the opposite could also happen, {\it i.e.}, that DM objects capture baryons.

These results are calculated for mirror DM, but are qualitatively valid for other kinds of DM that could form stable cores inside NS.
The basic prerequisites are that the DM particles are long-lived fermions with masses that are not too low or high compared to the baryon mass, and that they accumulate in sufficient number.
Obviously the details of the effect on the equilibrium sequences depend on the particular DM model used, but the trend is general and leads to more compact objects in presence of DM.
For example, it could be possible that, with some DM candidate, the observations are interpreted with a different EOS and smaller amounts of DM.
Anyway, whatever the DM candidate that can accumulate inside NS, the effect of adding a DM core is a softening of the EOS, thus mimicking that of other exotica as, {\it e.g.}, quark matter.

For example, DM particles with different masses and interactions than those considered here would affect the equilibrium sequence for a certain EOS in a different way.
In general, whatever the DM candidate is, if it accumulates inside NS the effect is an apparent softening of the EOS, thus mimicking that of other exotica, {\it e.g.}, quark matter.

There is, however, a possibility to distinguish between this and other scenarios, since the amount of DM in NS is expected to vary and depend on the whole history of the star, especially on the environments from which it originates and in which it lives.
This will result in a spread of mass-radius measurements that cannot be interpreted with a unique equilibrium sequence.
It seems unlikely that novel phases of ordinary matter can explain such an effect.

Should our hypothesis be confirmed by future observations of NS, a new possibility in DM research will arise: it will be possible to constrain DM properties by the inhomogeneities derived using the estimates of DM inside NS.
In fact, the comparison of masses and radii, derived from the observations, with the models computed for different DM contents, let us estimate the amount of DM in the core of each NS.
This is the integrated result of the DM present at the formation of the progenitor star and the DM accreted during the whole life of the star (progenitor and NS).
Correlating these results with the positions and orbits of NS, it should be possible to obtain information on the distribution of DM in the galaxy.
Furthermore, a reliable DM candidate should then also explain how it can accumulate in sufficient number in NS.
Considering the present estimated density for galactic DM, it's clear that a large amount of it cannot be accumulated only during the lifetime of the star, if we consider a smooth spatial distribution of DM.
One or both of the following requirements should be verified: either the DM is present already during the formation process; or the density distribution of DM is highly non-homogeneous, with the possibility of events like mergers with compact astrophysical objects with stellar sizes made of DM.

In conclusion, the recent observational results, which indicate that some neutron stars are extraordinary compact, can be explained with a nuclear matter equation of state if one allows for the possibility that dark matter contributes significantly to the mass of some neutron stars.
Improved and additional observational results will provide a crucial test of our hypothesis.
If confirmed, observational results for the masses and radii of neutron stars will become a powerful tool for the determination of dark matter.



\begin{thebibliography}{0}

\bibitem{Sandin:2008db}
  F.~Sandin and P.~Ciarcelluti,
  Astropart.\ Phys.\  {\bf 32}, 278 (2009)
  [arXiv:0809.2942 [astro-ph]].

\bibitem{Roos:2010wb}
  M.~Roos,
  arXiv:1001.0316 [astro-ph.CO].

\bibitem{Feng:2010gw}
  J.~L.~Feng,
  arXiv:1003.0904 [astro-ph.CO].

\bibitem{Kouvaris:2010vv}
  C.~Kouvaris and P.~Tinyakov,
  arXiv:1004.0586 [astro-ph.GA].

\bibitem{deLavallaz:2010wp}
  A.~de Lavallaz and M.~Fairbairn,
  arXiv:1004.0629 [astro-ph.GA].

\bibitem{Ozel:2006km}
  F.~Ozel,
  Nature {\bf 441}, 1115 (2006)
  [arXiv:astro-ph/0605106].

\bibitem{Ozel:2008kb}
  F.~Ozel, T.~Guver and D.~Psaltis,
  Astrophys.\ J.\  {\bf 693}, 1775 (2009)
  [arXiv:0810.1521 [astro-ph]].

\bibitem{Guver:2008gc}
  T.~Guver, F.~Ozel, A.~Cabrera-Lavers and P.~Wroblewski,
  Astrophys.\ J.\  {\bf 712}, 964 (2010)
  [arXiv:0811.3979 [astro-ph]].

\bibitem{Guver:2010td}
  T.~Guver, P.~Wroblewski, L.~Camarota and F.~Ozel,
  arXiv:1002.3825 [astro-ph.HE].

\bibitem{Ozel:2010fw}
  F.~Ozel, G.~Baym and T.~Guver,
  arXiv:1002.3153 [astro-ph.HE].

\bibitem{Drago:2010rp}
  A.~Drago and A.~Lavagno,
  arXiv:1004.0325 [astro-ph.SR].

\bibitem{Klahn:2006iw}
  T.~Klahn et al.,
  Phys.  Lett. B 654, 170 (2007)
  [arXiv:nucl-th/0609067].

\bibitem{Alford:2007}
  M.~Alford et al.,
  Nature, 445, E7 (2007)
  [arXiv:astro-ph/0606524].

\bibitem{Foot:1991bp}
  R.~Foot, H.~Lew and R.~R.~Volkas,
  Phys.\ Lett.\  B {\bf 272} (1991) 67.

\bibitem{Foot:2004pa}
  R.~Foot,
  Int.\ J.\ Mod.\ Phys.\  D {\bf 13}, 2161 (2004)
  [arXiv:astro-ph/0407623].

\bibitem{Ciarcelluti:2003wm}
  P.~Ciarcelluti,
  arXiv:astro-ph/0312607.

\bibitem{Ciarcelluti:2009da}
  P.~Ciarcelluti,
  arXiv:0911.3592 [astro-ph.CO].

\end{thebibliography}

\end{document}